# Broken mirror symmetry in excitonic response of reconstructed domains in twisted MoSe$_2$/MoSe$_2$ bilayers

Jiho Sung[1,2], You Zhou[1,2], Giovanni Scuri[2], Viktor Zólyomi[3], Trond I. Andersen[2], Hyobin Yoo[2], Dominik S. Wild[2], Andrew Y. Joe[2], Ryan J. Gelly[2], Hoseok Heo[1,2], Damien Bérubé[4], Andrés M. Mier Valdivia[5], Takashi Taniguchi[6], Kenji Watanabe[6], Mikhail D. Lukin[2], Philip Kim[2,5], Vladimir I. Fal'ko[7,8†] & Hongkun Park[1,2†]

[1]Department of Chemistry and Chemical Biology and [2]Department of Physics, Harvard University, Cambridge, MA 02138, USA

[3]Hartree Centre, STFC Daresbury Laboratory, Daresbury, WA4 4AD, United Kingdom

[4]Department of Physics, California Institute of Technology, Pasadena, CA, 91125, USA

[5]John A. Paulson School of Engineering and Applied Sciences, Harvard University, Cambridge, MA 02138, USA

[6]National Institute for Materials Science, 1-1 Namiki, Tsukuba 305-0044, Japan

[7]National Graphene Institute, University of Manchester. Booth St. E., Manchester M13 9PL, United Kingdom

[8]Henry Royce Institute for Advanced Materials, University of Manchester, Manchester M13 9PL, United Kingdom

†To whom correspondence should be addressed: Hongkun_Park@harvard.edu and Vladimir.Falko@manchester.ac.uk

**Structural engineering of van der Waals heterostructures via stacking and twisting has recently been used to create moiré superlattices[1,2], enabling the realization of new optical and electronic properties in solid-state systems. In particular, moiré lattices in twisted bilayers of transition metal dichalcogenides (TMDs) have been shown to lead to exciton trapping[3-8], host Mott insulating and superconducting states[9], and act as unique Hubbard systems[10-13] whose correlated electronic states can be detected and manipulated optically. Structurally, these twisted heterostructures also feature atomic reconstruction and domain formation[14-20]. Unfortunately, due to the nanoscale sizes (~10 nm) of typical moiré domains, the effects of atomic reconstruction on the electronic and excitonic properties of these heterostructures could not be investigated systematically and have often been ignored. Here, we use near-0° twist angle MoSe$_2$/MoSe$_2$ bilayers with large rhombohedral AB/BA domains[21] to directly probe excitonic properties of individual domains with far-field optics. We show that this system features broken mirror/inversion symmetry, with the AB and BA domains supporting interlayer excitons with out-of-plane (z) electric dipole moments in opposite directions. The dipole orientation of ground-state Γ-K interlayer excitons ($X_{I,1}$) can be flipped with electric fields, while higher-energy K-K interlayer excitons ($X_{I,2}$) undergo field-asymmetric hybridization with intralayer K-K excitons ($X_0$). Our study reveals the profound impacts of crystal symmetry on TMD excitons and points to new avenues for realizing topologically nontrivial systems[22,23], exotic metasurfaces[24,25], collective excitonic phases[26-28], and quantum emitter arrays[29,30] via domain-pattern engineering.**

To date, most studies of twisted TMD bilayers have assumed a rigid rotation of layers without atomic-scale rearrangement[3-8]. However, recent theoretical[14-16] and experimental[17-20] studies have amply demonstrated that even in these van der Waals heterostructures, the interlayer interactions can cause significant lattice reconstruction and the resultant domain formation. Understanding how these local atomic rearrangements impact the electronic and excitonic properties of TMD heterostructures is crucial in harnessing the full potential of the so-called moiré engineering.

Here, we study, for the first time, spatially resolved spectroscopic properties of distinct TMD bilayer domains using twisted $MoSe_2$ homo-bilayers (t-$MoSe_2$/$MoSe_2$) as a model system. We fabricate devices incorporating t-$MoSe_2$/$MoSe_2$ encapsulated by hexagonal boron nitride (hBN) using the 'tear-and-stack' technique[31]. Figure 1a shows a dark-field transmission electron microscopy (TEM) image of a near-0° t-$MoSe_2$/$MoSe_2$ device, **D1**, showing black and grey regions that correspond to alternating domains with rhombohedral stacking symmetry[21] (structures schematically shown in Fig. 1b). This domain formation is caused by the rearrangement of atoms within each TMD layer to preserve interlayer commensurability, similar to those observed in twisted bilayer graphene[32] and graphene on hBN[33]. Interestingly, for near-0° target twist angle, the regime we call a 'marginal twist', we observe irregular, micron-sized AB ($Mo_{top}Se_{bottom}$) and BA ($Se_{top}Mo_{bottom}$) domains that are large enough to be imaged optically. The irregularity of these domains likely reflects locally varying twist angles caused by strain inhomogeneity[34].

For t-$MoSe_2$/$MoSe_2$ devices used for optoelectronic characterizations, we include top and bottom graphene gates to independently control the out-of-plane electric field ($E_z$) and electrostatic doping (Fig. 1c)[35,36]. Figure 1d shows an integrated photoluminescence (PL) intensity map from a device **D2** under 1.88 eV (660 nm) excitation at 4 K. Compared to the bright, momentum-direct intralayer exciton emission from monolayer regions[37,38], the PL intensity of the twisted bilayer is reduced by three orders of magnitude, suggesting a direct-to-indirect band-gap transition from a monolayer to a bilayer[39-41]. The PL spectra of the twisted bilayer exhibit high energy peaks near 1.6 eV, which we label as $X_0$, and low energy peaks around 1.4 eV denoted as $X_{I,1}$ (Fig. 1e, see also Figs. S1 and S2 for the discussion of the multiple peaks). Similar to the case of other TMD bilayers[35,42-44], we assign the $X_0$ peaks to momentum-direct intralayer excitons composed of an electron and a hole residing in the same layer and the $X_{I,1}$ peaks to momentum-indirect interlayer excitons in which an electron and a hole are in separate layers (see below).

Figure 2a shows PL spectra of the $X_{I,1}$ peaks as a function of $E_z$ collected from spot 1 in device **D2**. The $X_{I,1}$ peaks shift linearly with $E_z$ by as much as 50 meV, indicating that the interlayer excitons responsible for them possess a finite electric dipole moment in the out-of-plane direction. From the slope of the field-dependent energy shift, we can obtain the electron-hole distance in the $X_{I,1}$-interlayer exciton using the formula $\Delta E = -edE_z$. Here $E$ is the energy of emission, $e$ is the elementary charge, and $d$ is the distance between the electron and the hole. The value of $d$ extracted from the data in Fig. 2a is 0.26 nm, which is smaller than the interlayer distance of ~0.6 nm[42,45] (See Supplementary Fig. 1). Interestingly, when $E_z$ falls below -0.09 V/nm, the slope of the linear Stark effect abruptly changes sign, indicating that the dipole moment direction changes at that field.

Importantly, at $E_z = 0$, the sign (but not the magnitude) of the Stark shift varies from spot to spot: for instance, the $X_{I,1}$ peaks at spot 2 in **D2** exhibit Stark shifts with an opposite slope from those at spot 1 (Figs. 2a and 2b). Spot 3, on the other hand, exhibits Stark shifts with both positive and negative slopes (Fig. 2c). These observations indicate that while the dipole-moment magnitude of the $X_{I,1}$-interlayer exciton is constant everywhere, its direction flips from spot to spot across the device, with the negative (positive) Stark slope signifying the dipole moment pointing up (down). Such behavior is unique to near-0° t-MoSe$_2$/MoSe$_2$ devices: for instance, in MoSe$_2$ devices that incorporate a natural (untwisted) 2H bilayer (Supplementary Fig. 2), we observe the same Stark shifts at all locations.

To generate a map of the $X_{I,1}$ dipole orientation across **D2**, we measure the integrated PL intensity below 1.36eV at $E_z = \pm 0.15$ V/nm (designated as PL+ and PL-, respectively) and calculate the ratio $\eta = \frac{PL_+ - PL_-}{PL_+ + PL_-}$: a positive (negative) η value indicates that the preferred dipole orientation is up (down). The η map in Fig. 2d clearly shows spatial domains of up (red) and down (blue) $X_{I,1}$ dipole orientations. In locations noted as white, Stark shifts of both positive and negative slopes appear, likely because the domain sizes are smaller than the probe beam size (~500 nm) and we are collecting PL spectra from both AB/BA domains.

Figures 3a-c show reflectance spectra of the $X_0$ peaks (~1.6 eV) as a function of $E_z$ at spots 1-3 in **D2** indicated in Fig. 2d (see Supplementary Fig. 3 for PL spectra in the same energy range). The energies of the two main resonances at 1.594 eV and 1.610 eV do not vary at small $E_z$, indicating that the $X_0$ peaks originate from intralayer excitons with essentially zero out-of-plane electric dipole moment[36,46,47]. At large $|E_z|$, however, the $X_0$ exciton peaks exhibit avoided crossings with a new spectral feature $X_{I,2}$, which exhibits a large Stark shift (black arrows in Fig. 3a)[13]. The observed slope from the avoided crossing indicate an electric dipole moment value that corresponds to an electron-hole separation of 0.63 nm (Figs. 3a-c).

Similar to the behavior of the $X_{I,1}$ peaks observed in Figs. 2a-c, the field-dependent avoided crossings between the $X_0$ and $X_{I,2}$ peaks also exhibit spatial variation across the device. As shown in Fig. 3, the avoided crossing for the lower $X_0$ peak occurs at $E_z = 0.07$ V/nm at spot 1 (Fig. 3a), at $E_z = -0.07$ V/nm at spot 2 (Fig. 3b), and for both polarities at spot 3 (Fig. 3c). Importantly, the map of avoided crossing patterns (Supplementary Fig. 4) agrees well with the dipole orientation map in Fig. 2d, strongly suggesting a common physical origin.

The spatially dependent optical properties in Figs. 2 and 3 can be understood from the electronic band structure of AB/BA ($Mo_{top}Se_{bottom}/Se_{top}Mo_{bottom}$) domains in t-MoSe$_2$/MoSe$_2$. Unlike natural 2H MoSe$_2$ bilayers, the crystal structures of the AB (BA) domains are not mirror/inversion symmetric: in particular, because the top and bottom layers are clearly distinguished by the crystal structure, electrons and holes can preferentially reside in the top or bottom layers, leading to interlayer excitons with preferred dipole orientation.

The broken mirror symmetry of the AB ($Mo_{top}Se_{bottom}$) domain is reflected in the properties of the states at its band edges[48,49]. Density functional theory (DFT) calculations of the AB-stacked MoSe$_2$/MoSe$_2$ bilayer show that the valence band maximum (VBM) is at the $\Gamma$ point, and the conduction band minimum (CBM) can be at the Q or K point depending on calculation parameters (Fig. 4a and Supplementary Fig. 5). While the hole wavefunction at the $\Gamma$ point is equally distributed over both layers, the electron wavefunction is more localized in the top layer (as compared to the bottom) regardless of whether the CBM is at the Q or K point (Fig. 4b). We note that this asymmetry is much stronger for the K-point band extremum (100%) than for the Q-point

(~60%). These results indicate that electrons prefer to reside in the top layer in AB-stacked MoSe$_2$/MoSe$_2$, and consequently the momentum-indirect interlayer excitons responsible for the X$_{I,1}$ peaks should have a downward dipole orientation. Because BA stacking is just a mirror image of AB stacking with respect to the horizontal plane, the preferred dipole orientation of the X$_{I,1}$-interlayer exciton is upwards in BA domains.

Quantitative comparisons between experimental data and DFT calculations provide further insight into the nature of the X$_{I,1}$-interlayer excitons. The DFT calculations show that electron-hole separations for Γ–K and Γ–Q interlayer excitons are 0.34 nm and 0.07 nm, respectively (Fig. 4a). Experimentally, the electron-hole separation deduced from the X$_{I,1}$-peak Stark shifts (Figs. 2a-c) is 0.26 nm. The comparison of dipole-moment values suggests that the X$_{I,1}$-interlayer excitons are likely from the Γ–K transition in AB/BA domains of t-MoSe$_2$/MoSe$_2$[43]. The field-dependent flip of dipole orientation observed in Figs. 2a-b provides further support for this assignment: by extrapolating the line with a negative slope at spot 2 to $E_z = 0$, we find a zero-field splitting of 43 meV between the two dipole orientations, which agrees well with DFT calculations for the two Γ–K transitions (the energy separations between two Γ–Q transitions should be orders of magnitude larger: see Supplementary Fig. 6).

The avoided crossing behavior between the X$_0$ and X$_{I,2}$ peaks in Figs. 3a-c can also be understood from DFT calculations of AB-stacked MoSe$_2$/MoSe$_2$. In AB-stacked MoSe$_2$/MoSe$_2$, the two monolayers exhibit staggered band alignment at the K (or K') point, and the K-K intralayer excitons (X$_0$) in the bottom layer have a lower energy than those in the top layer (Fig. 4c): this explains the origin of the two $X_0$ peaks observed in Fig. 1e and Figs. 3a-c. Meanwhile, the electron and the hole from different layers can form momentum-direct, K-K interlayer excitons as shown in Fig. 4c[13,50-55]. These K-K interlayer excitons responsible for $X_{I,2}$ peaks do not show any reflection contrast at $E_z=0$ but acquire oscillator strength and cause avoided crossings when they become resonant with the $X_0$ intralayer excitons, as shown in Fig. 4d[13]. The avoided crossing occurs at opposite $E_z$ in the AB and BA domains (Figs. 3a-c) because the preferred dipole orientation of the $X_{I,2}$ excitons is flipped in the two domains. Moreover, because the electrons and holes of the $X_{I,2}$ excitons are largely localized in separate MoSe$_2$ layers (Fig. 4b), the electron-hole separation should be equal to the interlayer distance (~0.6 nm), consistent with the separation extracted from the data in Figs. 3a-c. While further studies will be necessary to account for the coupling strengths between inter- and intralayer excitons[56-58], the band-structure presented in Fig. 4 provide a qualitative understanding of the main features in Figs. 2 and 3.

The domain-resolved spectroscopy of rhombohedral AB/BA domains in t-MoSe$_2$/MoSe$_2$ bilayers presented here demonstrates that local atomic registry and crystal symmetry have profound impacts on the exciton properties of these heterostructures. Specifically, our observations indicate that the interlayer excitons in AB and BA domains exhibit opposite electric dipole orientations dictated by crystal symmetry, as manifested by the spatial dependence of the $X_{I,1}$ Stark shift and the $X_0$-$X_{I,2}$ avoided crossing. Our observations indicate that the domain-specific, electrically tunable, optical properties of twisted TMD bilayers can be used to realize a wide variety of exciting potential applications. These include proposed excitonic topological insulators[22,23], quantum metasurfaces[24,25], and strongly correlated exciton lattices for Hubbard model physics[26-30]. Importantly, a tessellation of AB/BA domains in twisted TMD homo-bilayers[14-20] can be used to generate alternating dipolar exciton arrays whose relative energy and coupling strengths can be engineered by changing the twist angle and out-of-plane electric field. In such systems, it should also be possible to change the exciton-exciton interactions from attractive to repulsive[59] via field-dependent dipole flipping. For these reasons, twisted TMD bilayers with reconstructed lattices should open the door for the realization of many exotic exciton states, such as antiferroelectric exciton droplets[60], exciton liquids[61], and exciton condensates[62,63].

**Acknowledgments**

We thank Bernhard Urbaszek for helpful discussions. We acknowledge support from the DoD Vannevar Bush Faculty Fellowship (N00014-16-1-2825 for H.P., N00014-18-1-2877 for P.K.), NSF (PHY-1506284 for H.P. and M.D.L.), NSF CUA (PHY-1125846 for H.P. and M.D.L.), AFOSR MURI (FA9550-17-1-0002), ARL (W911NF1520067 for H.P. and M.D.L.), the Gordon and Betty Moore Foundation (GBMF4543 for P.K.), ONR MURI (N00014-15-1-2761 for P.K.), and Samsung Electronics (for P.K. and H.P.). V.I.F. acknowledges EPSRC grants No. EP/S019367/1, EP/S030719/1, EP/N010345/1, ERC Synergy Grant Hetero2D, Lloyd's Register Foundation Nanotechnology Grant, European Graphene Flagship Project and EC Project 2D-SIPC. The device fabrication was carried out at the Harvard Center for Nanoscale Systems. K.W. and T.T. acknowledge support from the Elemental Strategy Initiative conducted by the MEXT, Japan and the CREST (JPMJCR15F3), JST. D.B. acknowledges support from the Summer Undergraduate Research Fellowship at Caltech.

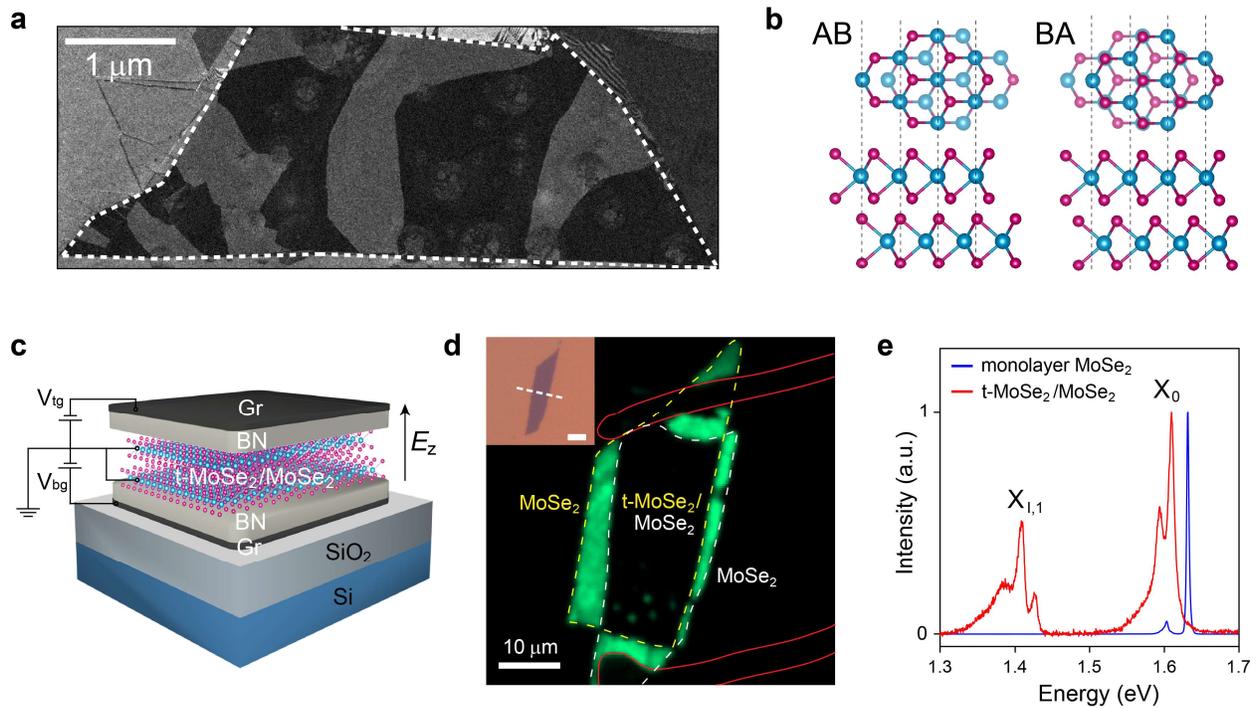

**Figure 1. AB/BA domains in t-MoSe$_2$/MoSe$_2$ devices. (a)** Dark-field TEM image of **D1** on a silicon nitride membrane obtained by filtering out all but one MoSe$_2$ diffraction peak ($g=10\bar{1}0$) with the device tilted off the zone axis. The twisted bilayer region (enclosed by the dashed white line) shows alternating, micron-sized, AB and BA domains. **(b)** Top and side views of atomic structures in twinned rhombohedral AB (Mo$_{top}$Se$_{bottom}$) and BA (Se$_{top}$Mo$_{bottom}$) stacking configurations. **(c)** Schematic of a t-MoSe$_2$/MoSe$_2$ device used for optoelectronic characterization. The out-of-plane electric field ($E_z$) and the doping concentration can be independently controlled by dual graphene gates. The black arrow defines the positive direction of $E_z$. **(d)** A PL map of **D2** at 4 K with 660-nm laser excitation. The monolayer region is enclosed by the yellow and white dashed lines and the overlapped region is t-MoSe$_2$/MoSe$_2$. The solid red lines show the outline of a Pt contact to the top and bottom MoSe$_2$ monolayer. The original MoSe$_2$ monolayer was torn along the dashed line indicated in the inset to generate t-MoSe$_2$/MoSe$_2$. Scale bar, 10 μm. **(e)** Normalized PL spectra from the t-MoSe$_2$/MoSe$_2$ (red) and monolayer MoSe$_2$ (blue) at $E_z = 0$.

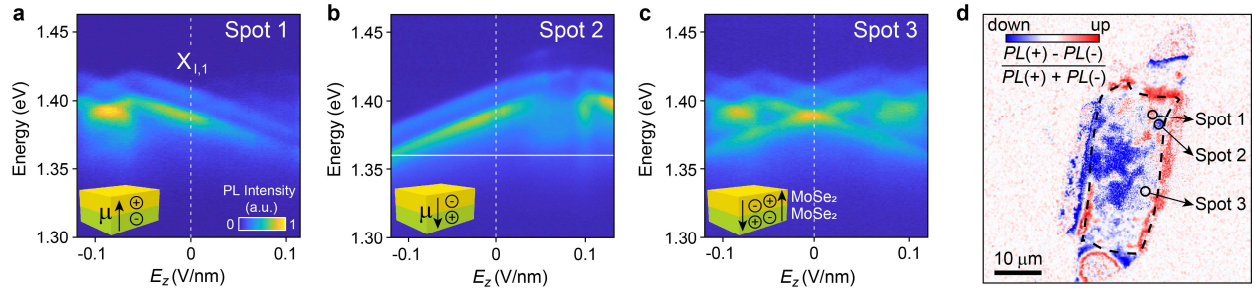

**Figure 2. Electric-field dependent PL spectra of the $X_{I,1}$ peaks obtained from D2 at 4K. (a-c)** PL spectra of the $X_{I,1}$ peaks as a function of the electric field at spot 1, 2, and 3, respectively. Inset: schematic of t-MoSe$_2$/MoSe$_2$, showing interlayer excitons with the preferred dipole orientations. White dotted lines, $E_z$ = 0 V/nm. **(d)** A map of η, defined by (PL(+) – PL(-))/(PL(+) + PL(-)). PL(±) is the PL intensity at $E_z$ = ±0.15 V/nm, integrated over the energy range below 1.36 eV (white solid line in **(b)**). Red, preferred dipole orientation up; blue, preferred dipole orientation down. The t-MoSe$_2$/MoSe$_2$ region is indicated by the dashed black line. Black solid circles represent spot 1, 2, and 3 where the spectra are taken.

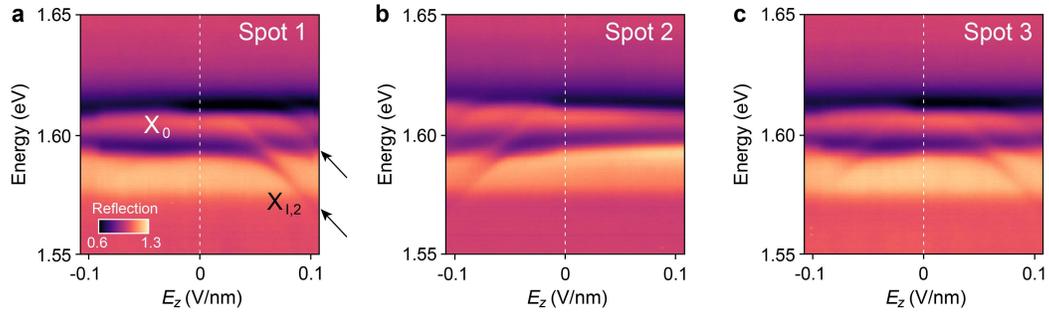

**Figure 3. Electric-field dependent reflectance spectra of the $X_0$ peaks obtained from D2 at 4K. (a-c)** Reflectance spectra as a function of the electric field at spot 1,2, and 3 indicated in Fig. 2d, respectively. White dotted lines, $E_z = 0$ V/nm. The $X_0$ excitons exhibit avoided crossings with $X_{I,2}$ excitons at **(a)** positive, **(b)** negative and **(c)** both polarities of $E_z$. We normalize the reflected intensity using the measured reflection in the same spot in the highly electron doped regime.

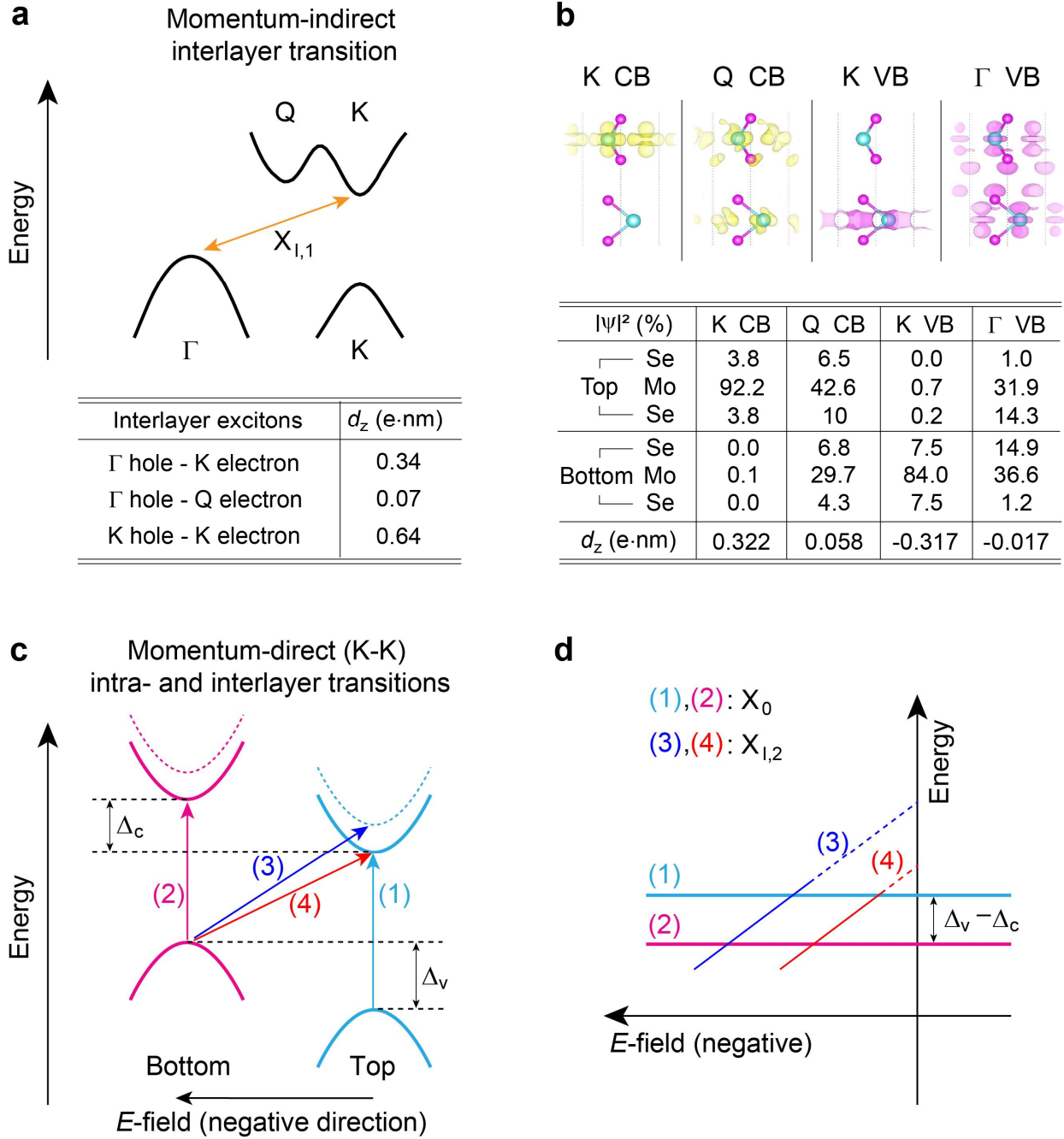

**Figure 4. Electronic band structure of AB-stacked MoSe$_2$/MoSe$_2$. (a)** Top: electronic band structure of an AB-stacked MoSe$_2$/MoSe$_2$ bilayer with the VBM at the Γ point and the CBM at either the K or Q points. The data in Fig. 2 suggest that momentum-indirect interlayer excitons, $X_{I,1}$ are from the Γ–K transition (Orange arrow). Bottom: the calculated dipole moment of interlayer excitons from momentum-indirect Γ–K and Γ–Q and momentum-direct K-K transitions.

**(b)** Top: contour plots of the electron wavefunctions at the conduction-band K and Q points and the hole wavefunctions at the valence-band K and Γ points. Bottom: atomic decomposition of wavefunctions at the K and Q points in the conduction band and that at the K and Γ points in the valence band. The calculated dipole moment of each state is shown in the last row. **(c)** The electronic band structure of AB-stacked $MoSe_2/MoSe_2$ at the K valley. Electronic states localized in the top (bottom) are drawn in cyan (magenta) color, while the solid and dashed line represent spin up and down states. The optical transitions (1) and (2) correspond to momentum-direct intralayer transitions ($X_0$), while (3) and (4) represent the momentum-direct (K-K) interlayer transitions ($X_{I,2}$). The value of $\Delta_c$ obtained from the DFT calculations is 50 meV and $\Delta_v$ is 62 meV at $E_z = 0$. The black arrow at the bottom represents the negative direction of an out-of-plane electric field. The energy of the interlayer transition from the bottom layer VBM to the top layer CBM (3 and 4) is reduced under negative $E_z$. **(d)** The energy of $X_0$ (1 and 2 in **c**) and $X_{I,2}$ (3 and 4 in **c**) transitions as a function of $E_z$. Due to the reduced binding energy of $X_{I,2}$, the energy of $X_{I,2}$ at $E_z = 0$ is higher than $X_0$.

**Supplementary Information for "Broken mirror symmtery in excitonic response of reconstructed domains in twisted MoSe$_2$/MoSe$_2$ bilayers"**

Jiho Sung[1,2], You Zhou[1,2], Giovanni Scuri[2], Viktor Zólyomi[3], Trond I. Andersen[2], Hyobin Yoo[2], Dominik S. Wild[2], Andrew Y. Joe[2], Ryan J. Gelly[2], Hoseok Heo[1,2], Damien Bérubé[4], Andrés M. Mier Valdivia[5], Takashi Taniguchi[6], Kenji Watanabe[6], Mikhail D. Lukin[2], Philip Kim[2,5], Vladimir I. Fal'ko[7,8]† & Hongkun Park[1,2]†

[1]Department of Chemistry and Chemical Biology and [2]Department of Physics, Harvard University, Cambridge, MA 02138, USA

[3]Hartree Centre, STFC Daresbury Laboratory, Daresbury, WA4 4AD, United Kingdom

[4]Department of Physics, California Institute of Technology, Pasadena, CA, 91125, USA

[5]John A. Paulson School of Engineering and Applied Sciences, Harvard University, Cambridge, MA 02138, USA

[6]National Institute for Materials Science, 1-1 Namiki, Tsukuba 305-0044, Japan

[7]National Graphene Institute, University of Manchester. Booth St. E., Manchester M13 9PL, United Kingdom

[8]Henry Royce Institute for Advanced Materials, University of Manchester, Manchester M13 9PL, United Kingdom

†To whom correspondence should be addressed: Hongkun_Park@harvard.edu and Vladimir.Falko@manchester.ac.uk

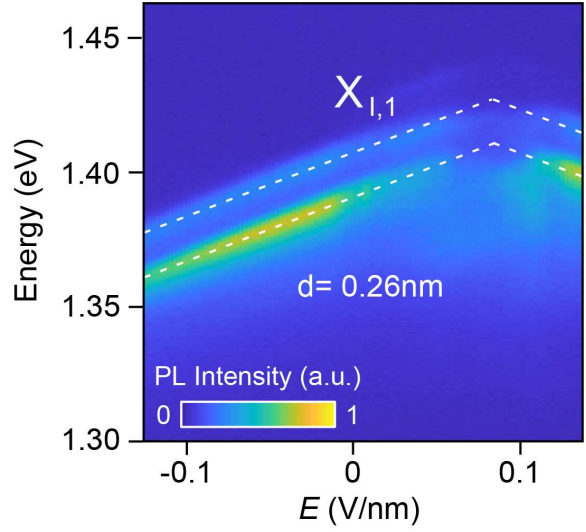

**Supplementary Figure 1. Electric dipole moment of the $X_{I,1}$-interlayer exciton and possible origin of the multiple peaks.** PL spectra of the $X_{I,1}$ peaks as a function of the electric field at spot 2 obtained from **D2**. The top and bottom gate voltages are swept together with a voltage ratio, -α ($\alpha = 39/42$, the bottom hBN thickness over the top hBN thickness, $V_{BG} = -\alpha \times V_{TG}$) in order to keep **D2** close to zero doping concentration, while applying an electric field. From the slope of the linear Stark shift, we obtain the magnitude of the interlayer exciton dipole moment using the formula $\Delta E = -edE_z$, where $E$ is the energy of emission, $e$ is the electric charge, $d$ is the distance between an electron and a hole, and $E_z$ is the applied out-of-plane electric field. The applied out-of-plane electric field was calculated with the formula, $E_z = (C_{Bottom,hBN}V_{BG} - C_{Top,hBN}V_{TG})/2\varepsilon_{biMoSe2}$, where $C_{Bottom,hBN} = \varepsilon_{hB}/d_{Bottom,hB}$ and $C_{Top,hBN} = \varepsilon_{hBN}/d_{Top,hBN}$ are the geometric capacitance of the bottom and top hBN, $V_{BG}$ and $V_{TG}$ are the applied bottom and top gate voltages and $\varepsilon_{biMoS}$ is the dielectric constant of bilayer MoSe$_2$. We use $\varepsilon_{hB} = 3.5$ and $\varepsilon_{biMoSe2} = 7.9$, which were reported in the literature [S1].

We observe multiple low energy PL peaks, separated by ~18 meV. The energy separation of these low energy PL peaks does not change as a function of the electric field. A possible origin of the multiple peaks is phonon replicas because the optical phonon energy in MoSe$_2$ is similar to the energy separation of the peaks. More experiments and theoretical studies are required to fully understand their origin. [S2, S3]

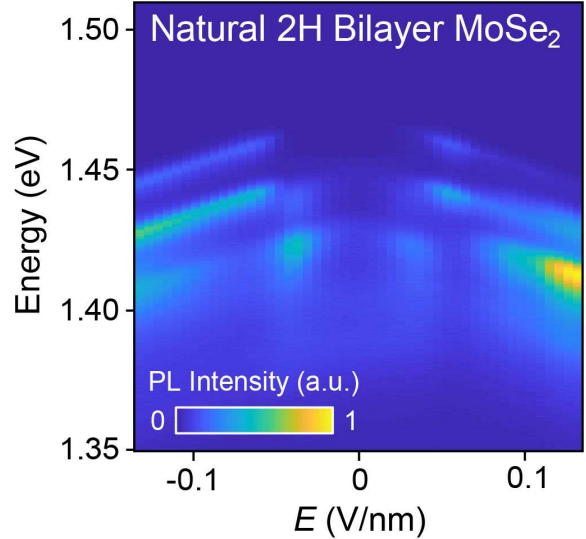

**Supplementary Figure 2. Electric-field dependent PL spectra in a natural bilayer MoSe$_2$ device, D3 at 4K.** The top and bottom gate voltages are swept together with a voltage ratio, -1 (the thicknesses of the top and bottom hBN are both 48 nm) in order to keep **D3** close to zero doping concentration. We observe three low energy PL peaks, separated by ~18 meV, similar to the multiple-peak structures near 1.4 eV in the t-MoSe$_2$/MoSe$_2$ bilayers. The energy separation between these low energy PL peaks does not change as a function of the electric field as well.

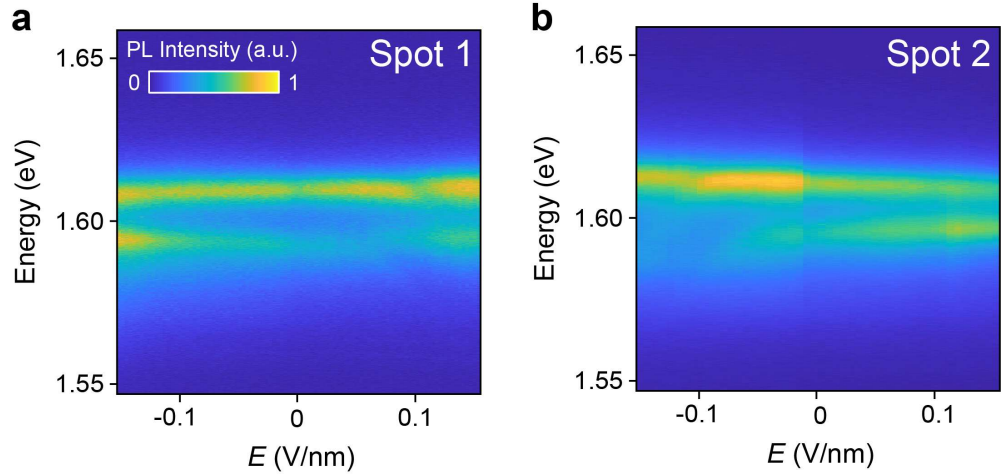

**Supplementary Figure 3. Electric-field dependent PL spectra of the $X_0$ peaks obtained from D2 at 4K.** We measure the PL spectra with the same gate operation scheme as the reflectance measurements in Fig. 3 in the main text at **(a)** spot 1 and **(b)** spot 2. Two strong PL peaks at 1.594 eV and 1.610 eV are observed, whose energies do not change as a function of the out-of-plane electric field. Similar to the reflectance measurements, we observe an avoided crossing for the lower $X_0$ peak near $|E_z|$ = 0.07 V/nm.

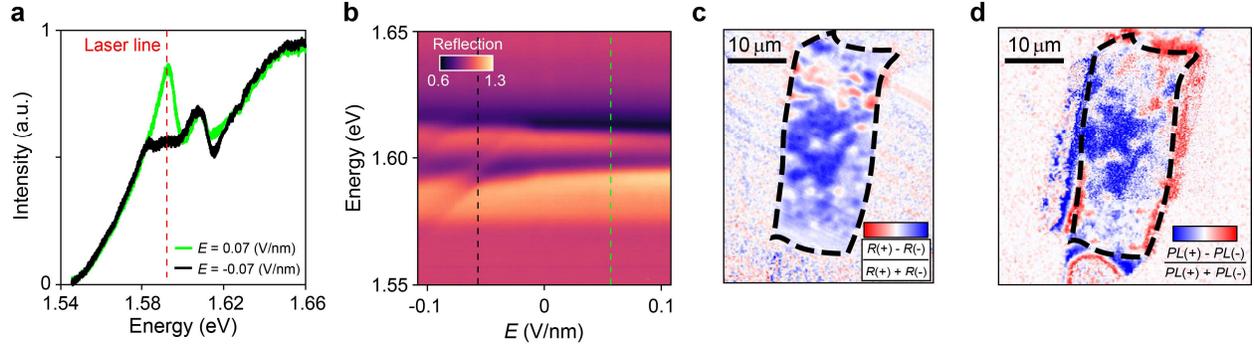

**Supplementary Figure 4. Spatial image of the avoided crossing features at 4K (a)** Cross sections of the reflectance spectra at spot 2 in **D2** along the green (at $E_z$ = 0.07 V/nm) and black (at $E_z$ = -0.07 V/nm) dashed lines in **(b)**. We spatially image the avoided crossing features by applying a constant electric field of opposite polarity (±0.07 V/nm) and map out the reflection with a continuous wave laser centered at 1.59 eV (red line in a) to spatially image the avoided crossing features. **(c)** A map of the ratio $\xi = \frac{R_+ - R_-}{R_+ + R_-}$, calculated from the raw reflection data is shown ($R_\pm$ is the reflected intensity at $E_z = \pm 0.07$ V/nm). The spatial dependence of the avoided crossing features matches well with the interlayer dipole orientation map in **(d)**, indicating that these features have the common physical origin. We flipped the color bar in (c) in order to match the color scheme with a map of $\eta = \frac{PL_+ - PL_-}{PL_+ + PL_-}$: this is because the reflected intensity is smaller when there is an avoided crossing.

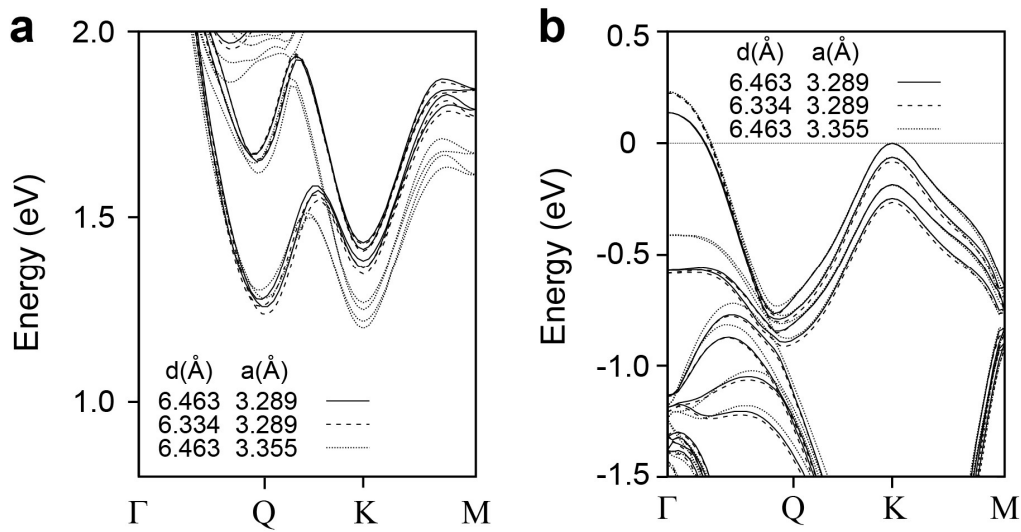

**Supplementary Figure 5. Electronic band structure of the AB-stacked MoSe$_2$/MoSe$_2$ bilayer from the DFT calculation.** DFT calculated **(a)** conduction and **(b)** valence bands of the AB-stacked MoSe$_2$/MoSe$_2$ bilayer, with the energies counted from the local valence band maximum (VBM) at K point. Interlayer distance, *d* and lattice constant, *a* are set to different values for which the band dispersions are plotted with solid, dashed and dotted lines.

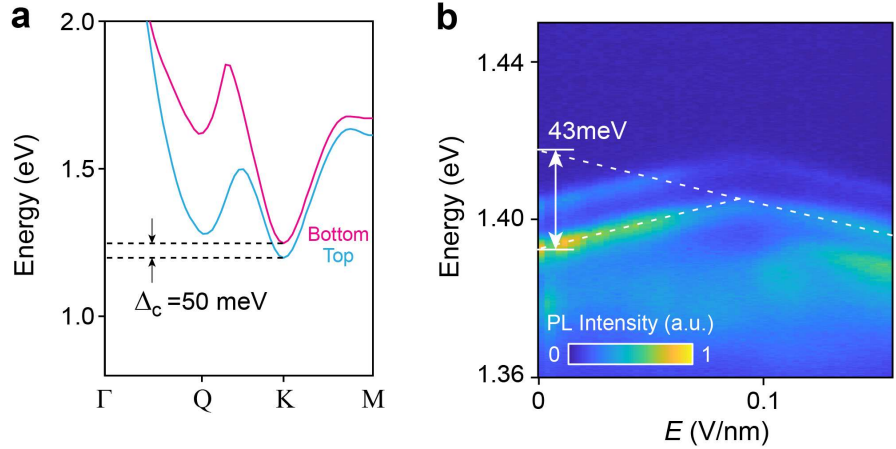

**Supplementary Figure 6. Estimating the splitting between the top and bottom layer K valley conduction band minimum (CBM). (a)** The calculated conduction band of AB-stacked MoSe$_2$/MoSe$_2$ using DFT. The band where electrons are more localized in the top (bottom) MoSe$_2$ layer is shown in cyan (magenta) color. At the K point, there is a splitting ($\Delta_c$) between the top and bottom MoSe$_2$ layer. At $E_z$=0, electrons in the K point, which is the CBM, are localized in the top layer. A positive electric field decreases $\Delta_c$ by making the electrons in the bottom layer more energetically favorable. Spin flipped upper conduction bands are not plotted for the clarity. The calculated value of $\Delta_c$ is 50 meV. **(b)** We can estimate $\Delta_c$ at zero electric field by extrapolating the line with a negative slope to zero field in the PL spectrum. The extracted value is ~43 meV, in good agreement with the calculated value of 50 meV. By comparison, the calculated conduction band splitting at the Q point is ~400 meV. This splitting would require a field larger than 3.3 V/nm to switch the preferred dipole orientation of the Γ–Q interlayer excitons (based on the calculated dipole moment of Q valley electrons of ~0.06 (e·nm)). This electric field value is much larger than the field at which we observe the transition (0.09 V/nm), suggesting that the CBM is at the K point.